\documentclass[journal]{IEEEtran}
\usepackage{cite}
\usepackage{amsmath,amssymb,amsfonts}
\usepackage{algorithmic}
\usepackage{graphicx}
\usepackage{textcomp}
\usepackage{xcolor}
\usepackage{cite}
\usepackage{amsmath,amssymb,amsfonts}
\usepackage{algorithmic}
\usepackage{graphicx}
\usepackage{textcomp}
\usepackage{xcolor}
\usepackage{subcaption}
\usepackage{multicol,multirow}
\usepackage{enumerate,textcomp,array}
\usepackage{placeins}
\usepackage{nicematrix}
\NiceMatrixOptions{cell-space-top-limit = 1pt,cell-space-bottom-limit = 1pt}
\usepackage{threeparttable}
\usepackage{dblfloatfix}
\usepackage{comment}
\usepackage{array}
\newcolumntype{?}{!{\vrule width 1pt}}
\usepackage{gensymb}

\usepackage{filecontents}
\usepackage{tikz, tikzscale, pgfplots}
\usetikzlibrary{calc}

\newcommand{\rmd}{\mathrm{d}}

\newcommand{\bA}{\mathbf{A}}

\newcommand{\mn}{\mathcal{N}}
\newcommand{\mg}{\mathcal{G}}
\newcommand{\mt}{\mathcal{T}}
\newcommand{\ms}{\mathcal{S}}
\newcommand{\ml}{\mathcal{L}}
\newcommand{\mr}{\mathcal{R}}

\usepackage{url}

\usepackage{breakurl}
\usepackage{booktabs}

\def\BibTeX{{\rm B\kern-.05em{\sc i\kern-.025em b}\kern-.08em
    T\kern-.1667em\lower.7ex\hbox{E}\kern-.125emX}}
\begin{document}

\title{Nodal Frequency Stability-Constrained UC \& ED for Renewable-Dominated Power Systems\\}

\author{Marena~Trujillo,~\IEEEmembership{Student Member,~IEEE,}
        David~Cole,
        and~Bri-Mathias~Hodge,~\IEEEmembership{Senior Member,~IEEE}
\thanks{This material is based upon work supported by the National Science Foundation Graduate Research Fellowship under Grant No. DGE $2040434$.}%
\thanks{M. Trujillo and B.M. Hodge are with the Department
of Electrical, Computer \& Energy Engineering and the Renewable and Sustainable Energy Institute, University of Colorado Boulder, Boulder,
CO, 80309 USA (e-mail: marena.trujillo@colorado.edu, brimathias.hodge@colorado.edu)}
\thanks{D. Cole is with the Andlinger Center for Energy and the Environment, Princeton University, New Jersey
08540, USA (e-mail: dc0173@princeton.edu).}
}

\maketitle
\thispagestyle{empty}
\pagestyle{plain}
\begin{abstract}
In modern power systems with high shares of renewable, inverter-based resources (IBRs), frequency stability becomes more complex due to the fast dynamics of IBRs and frequency trajectories that vary significantly from bus to bus. In this paper, we present an optimization framework for unit commitment and economic dispatch with endogenous frequency stability constraints at each bus. \textcolor{black}{Two approaches for mitigating excessively low instantaneous frequency values in the event of the largest generator contingency} are proposed: 1) by introducing a constraint requiring more thermal generation, and 2) by constraining the maximum power output of the generator that had the largest power output in the incumbent solution. Both approaches proved effective in eliminating dispatch scenarios that resulted in instantaneous frequencies below $58$ Hz, while the second approach minimized the difference in production cost values from the non-stability-constrained case. Overall, the results indicate that the proposed optimization framework is a more effective alternative to frequency stability-constrained unit commitment and economic dispatch (UC \& ED) than those based on the center-of-inertia (COI) principle.
\end{abstract}

\begin{IEEEkeywords}
Renewable-dominated systems, Grid-forming inverters, Frequency stability
\end{IEEEkeywords}
\section{Introduction}\label{intro}
The decarbonization of the electric power sector is among the most urgent priorities in global efforts to mitigate the worst effects of climate change \cite{calvinIPCC2023Climate2023}. As a result, the adoption of wind, solar, and battery storage devices, all of which are inverter-based resources (IBRs), has accelerated. Importantly, 
IBRs exhibit fundamentally different dynamic behavior than synchronous generators (SG) \cite{milanoFoundationsChallengesLowInertia2018}. \textcolor{black}{Of the different types of IBRs, we distinguish between grid-following (GFL) and grid-forming (GFM) inverters. As more IBRs are introduced into grid systems, GFMs are increasingly attractive for the ability to independently form frequency and voltage \cite{linResearchRoadmapGridForming}. Furthermore, GFMs may operate at much faster timescales than SGs because they are not inhibited by the inertia of a spinning rotor, rather, their dynamics are primarily a function of their control architectures \cite{kenyonInteractive}. As a consequence, the frequency response of a GFM can look quite different from the typical oscillatory response of an SG; however, the degree to which an inverter's frequency response differs from an SG depends on its specific control scheme. As the name suggests, virtual synchronous generators are a class of GFMs meant to mimic the behavior of an SG, while droop control GFMs and virtual oscillator control display faster dynamics \cite{linResearchRoadmapGridForming}}. These disparate device-level frequency dynamics extend to network-wide dynamics; the average frequency response of an inverter-dominated network is different from that of an SG-dominated network \cite{heAnalysisControlFrequency2024}. 

In renewable-dominated systems, the competing timescales of inverter and synchronous generator dynamics can result in erratic frequency deviations following a disturbance \cite{kenyonInteractive}. Increased localization of disturbance response is also observed in inverter-dominated networks \cite{baughman2026implicationsgridforminginverterparameters}. This rise in frequency heterogeneity is at odds with the common power system assumption of a global network frequency, at least directly following a disturbance. Faster frequency dynamics and increased localization make power systems more vulnerable to protective action caused by local violations of underfrequency or rate-of-change-of-frequency (RoCoF) limits \cite{heAnalysisControlFrequency2024,nahid-al-masoodLowInertiaPower2016}.

Evidence of renewable-dominated systems with increased nodal frequency heterogeneity has already been observed in large-scale systems. For example, low inertia and staunch load shedding limits contributed to the 2019 Great Britain power outage \cite{GBPowerSystem2020}. During this lightning strike-induced event, frequency dropped suddenly, triggering the RoCoF protection of some units and resulting in $500$ MW of distributed generation being tripped. Later on, $200$ MW of distributed generation tripped when the system frequency dipped below $49$ Hz \cite{bialekWhatDoesGB2020}. Additionally, undesirable IBR tripping due to low frequency occurred during an operating incident in Central Queensland in June 2021 \cite{AemocomauMediaFiles}. In response, the Australian Energy Market Operator (AEMO) initiated a review of the efficacy of underfrequency load shedding (UFLS) schemes during times of high IBR penetration \cite{AemocomauMediaFilesa}. Reports authored by the North American Electric Reliability Corporation (NERC) have also identified low inertia and unexpected IBR tripping during frequency events as operational challenges \cite{NerccomGlobalassetsOurwork,NerccomGlobalassetsPrograms}.

It is standard practice for detailed dynamic stability assessment to be distinct from unit commitment and economic dispatch (UC \& ED) processes. While standard reserve requirements are useful for ensuring system reliability following an unplanned outage, they are technology agnostic and only consider static security criteria, making them insufficient for ensuring frequency stability \cite{conejo2017unit}. However, it would be desirable to unify frequency-stability assessment and UC \& ED to improve the robustness of UC \& ED decisions in the face of emerging dynamic issues introduced by IBRs. While there have been numerous efforts to formulate frequency stability constraints for UC \& ED problems, many efforts are predicated on the center-of-inertia concept (COI) \cite{nguyenOptimalPowerFlow2019}, which is ill-suited for capturing the frequency heterogeneity of large, renewable-dominated power systems with significant levels of IBRs  \cite{kamwaAutomaticSegmentation,baughman2026implicationsgridforminginverterparameters}, in particular GFMs. At the same time, methodologies that allow for more complex evaluation of frequency stability require external power system simulation software, large datasets, or are computationally burdensome \cite{liuAnalyticalModelFrequency2020}. In this paper, a novel computationally tractable \textit{nodal} frequency-constrained UC \& ED formulation is proposed. This methodology allows for accurate and interpretable analytical frequency-stability constraints that are suitable for mixed-generation power systems.

The rest of the paper is structured as follows. A discussion of the literature related to frequency stability-constrained UC \& ED can be found in Section \ref{litreview}. In Section \ref{branchcut}, background information on branch-and-cut optimization is provided, followed by an overview of the frequency stability model and optimization problem in Section \ref{overview}. Results of the case study using a $73$-bus system are included in Section \ref{casestudy}, and the efficacy of multiple solution approaches is compared. Concluding thoughts are provided in Section \ref{conclusion}.
\section{Frequency Stability-Constrained Power Systems Optimization}\label{litreview}
There have been numerous efforts to incorporate stability constraints in power systems optimization problems like optimal power flow (OPF) and UC\&ED. However, accurately representing dynamic behavior within an optimization framework is challenging. One approach is to simplify the dynamics of an entire grid system by representing a network of machines as a single equivalent machine. This is the approach taken in \cite{nguyenOptimalPowerFlow2019} and \cite{badesaSimultaneousSchedulingMultiple2019}. In \cite{nguyenOptimalPowerFlow2019}, the authors propose a frequency security-constrained optimal power flow framework, however, the frequency behavior of the entire system is represented by an aggregated single machine model and does not take into consideration the heterogeneity of frequency response based on electrical distance. In this framework, the dynamics of IBRs are represented by altering the parameters of an SG model to reflect diminished inertia. Similarly, the stochastic unit commitment framework in \cite{badesaSimultaneousSchedulingMultiple2019} utilizes a uniform frequency model that assumes dominant SG dynamics. \textcolor{black}{Likewise, frequency response is modeled using the COI concept in \cite{pyschel-lovengreenFrequencyResponseConstrained2018}. Here, the authors acknowledge the key role of contingency size on frequency excursion magnitude \cite{pyschel-lovengreenFrequencyResponseConstrained2018}. In Wang et al. \cite{wang2025nodal}}, the authors use a hybrid data-model-driven approach to generate the
nodal frequency constraints for an energy storage optimization problem, but the approach requires first carrying out a full-year COI frequency-constrained unit commitment (UC). 


Another approach is to model the dynamics of  individual generating devices and relate them using algebraic network equations. The result are systems of differential-algebraic equations (DAEs), which must be added as constraints to an optimization problem. DAE constraints in transient stability constrained optimal power flow problems have been handled using the following broad categories of techniques \cite{zhangAdvancesTransientStabilityconstrained2025}:
1) sequential methods, 2) discretization methods, 3) heuristic algorithms, 4) machine learning-based approaches. In the literature, discretization methods are most frequently utilized. For instance, in \cite{ganStabilityconstrainedOptimalPower2000}, dynamic equations are converted to algebraic equations and added to a standard OPF framework. The objective function and constraints are linearized. In that work, the rotor angle is used to determine the stability of an SG-only system. Note that in SG-dominated systems, rotor angle instability is the first indicator of instability, but this may not be the case for systems with high shares of IBRs.

A framework for frequency stability-constrained optimal power flow is detailed in \cite{zhaoFrequencyStabilityConstrained2021}. The DAEs describing the frequency dynamics of SGs are converted to numerically equivalent algebraic equations. A re-dispatch method is used to ensure security during primary frequency response, and a constraint on nadir is implemented. While the authors include a general discussion on the discretization of differential algebraic constraints, the relevant constraints for IBRs are not explicitly presented. For large systems with many generators, the computational burden increases significantly. 
\textcolor{black}{In Kilembe and Papadopolous \cite{kilembeRegulationDisturbanceMagnitude2023}, machine learning techniques are use to capture local frequency dynamics rather than the COI frequency, and a disturbance size regulation scheme for limiting frequency nadir and RoCoF is proposed. However, many simulations must be carried out to obtain the data necessary for training the model, and a physical interpretation of the hyperparameters is not provided.}


While COI-based approaches are attractive for their low computational burden, the COI frequency of a system is ill-suited for representing the frequency dynamics of large systems with high shares of IBRs, where there may be numerous coherent areas and increased frequency localization due to fast IBR dynamics \cite{baughman2026implicationsgridforminginverterparameters}. The COI frequency is also of limited value for mitigating generator tripping due to instantaneous nadir levels due to the inherent heterogeneous nature of frequency response across a system. In other words, while the COI frequency provides valuable information about the average frequency response across a system, it does not provide insight into the location and magnitude of the most severe frequency deviations in a system \cite{Anderson1990ALS}, whereas protection devices use only local information.

This paper presents a novel nodal frequency stability-constrained UC \& ED optimization framework. Unlike the vast majority of the literature on frequency stability-constrained UC \& ED, this framework assesses frequency stability at the nodal level, rather than by coherent areas or by assuming a COI frequency. \textcolor{black}{Furthermore, analytical models with straightforward physical interpretations are utilized for frequency stability assessment.} This detailed approach enables operators to arrive at commitment decisions that are less likely to result in IBR tripping caused by instantaneous frequency deviations.
\section{Branch and Cut Method}\label{branchcut}
An overview of branch-and-cut optimization is provided in this section, followed by a discussion on the application of this approach to the optimization problem at hand.

\subsection{Background}
Branch-and-cut algorithms have a long history in combinatorial optimization and can be utilized to solve a wide variety of problems \cite{mitchell2002branch,lucena1996branch,karamanov2006branch}. They are commonly applied to mixed-integer linear programs (MILPs), which are NP-hard and far more challenging to solve than linear programs. The branch-and-cut approach is the combination of two popular solution methods for MILP problems: the branch-and-bound algorithm and the cutting-plane algorithm \cite{karamanov2006branch}. Branch-and-bound algorithms prune regions of a problem's search tree that cannot lead to better solutions. The tree search strategy, branch strategy, and pruning rules may be fine-tuned to improve performance \cite{morrison2016branch}. In contrast, cutting-plane algorithms use inequalities to tighten problem formulations \cite{karamanov2006branch}. The powerful combination of these methods, called the branch-and-cut approach, involves generating cutting planes at the nodes of the branching tree to improve the lower bound \cite{karamanov2006branch}.

\subsection{Application to UC \& ED Problem}\label{sec:apptoUCED}
The optimization problem considered in this paper is a MILP with additional scenario-dependent feasibility requirements arising from frequency stability considerations. Directly embedding detailed frequency dynamics and nodal instantaneous frequency constraints into the UC \& ED formulation would result in a prohibitively large and computationally intractable problem for large-scale power systems. To address this challenge, a branch-and-cut approach is employed, allowing frequency stability constraints to be enforced implicitly.

In the proposed branch-and-cut framework the optimization problem is solved without any modeling of frequency stability. At each new incumbent solution of the MILP, the solution of each of the first 24 hrs of the period is passed to the frequency response model described in Section IV under the largest generator contingency where the generator with the largest power output at that time period is tripped. If the contingency results in any instantaneous nodal frequencies that violate the minimum allowable threshold, the solution at that hour is declared infeasible and a cutting plane is generated to eliminate it from the feasible region of the optimization problem. An additional constraint to the cutting plane is also added to further encourage feasible solutions. Two different constraint formulations are tested: first, a constraint requiring additional thermal generator commitment and second, a constraint on the maximum power output of the generator with the largest power output in the incumbent solution.

The cuts are added dynamically via solver callbacks in Gurobi during the branch-and-bound process, ensuring that only those regions of the solution space that are infeasible from a frequency stability perspective are removed. This iterative process continues until a solution is found that satisfies both the traditional UC \& ED constraints and the post-contingency frequency security requirements. Code for implementing this approach is available at our public repository \cite{GithubRepo}. 
\section{Problem Overview}\label{overview}
\subsection{Frequency Response Model}
The frequency response model from \cite{Trujillo2025} is leveraged for finding stability-constrained solutions. In this simplified model, the network representation is based on the Kron-reduced DC power flow equation, where generator nodes are retained \cite{trujilloAnalyticalModelsFrequency2026}. For a set of buses, $\mn$, let $\gamma$ be the set of generator nodes, and $\beta$ be the set of all nodes without generators ($\beta = \mn \setminus \gamma$). To describe a system of $|\gamma|$ generation nodes, we define the following vector quantities:
\begin{equation}
\begin{split}
    \Delta \mathbf{p}_\Gamma &= \begin{bmatrix} \Delta p_{\gamma_1} \cdots \, \Delta p_{\gamma_{|\gamma|}} \end{bmatrix}^T\\
    \Delta \mathbf{\theta}_\Gamma &= \begin{bmatrix} \Delta \theta_{\gamma_1} \cdots \, \Delta \theta_{\gamma_{|\gamma|}} \end{bmatrix}^T\\
    \Delta \mathbf{p}_D &= \begin{bmatrix} \Delta p_{d_1} \cdots \, \Delta p_{d_{|\gamma|}} \end{bmatrix}^T.
\end{split}
\end{equation}

\begin{figure}[htpb]
    \centering
    \includegraphics[width=.6\linewidth]{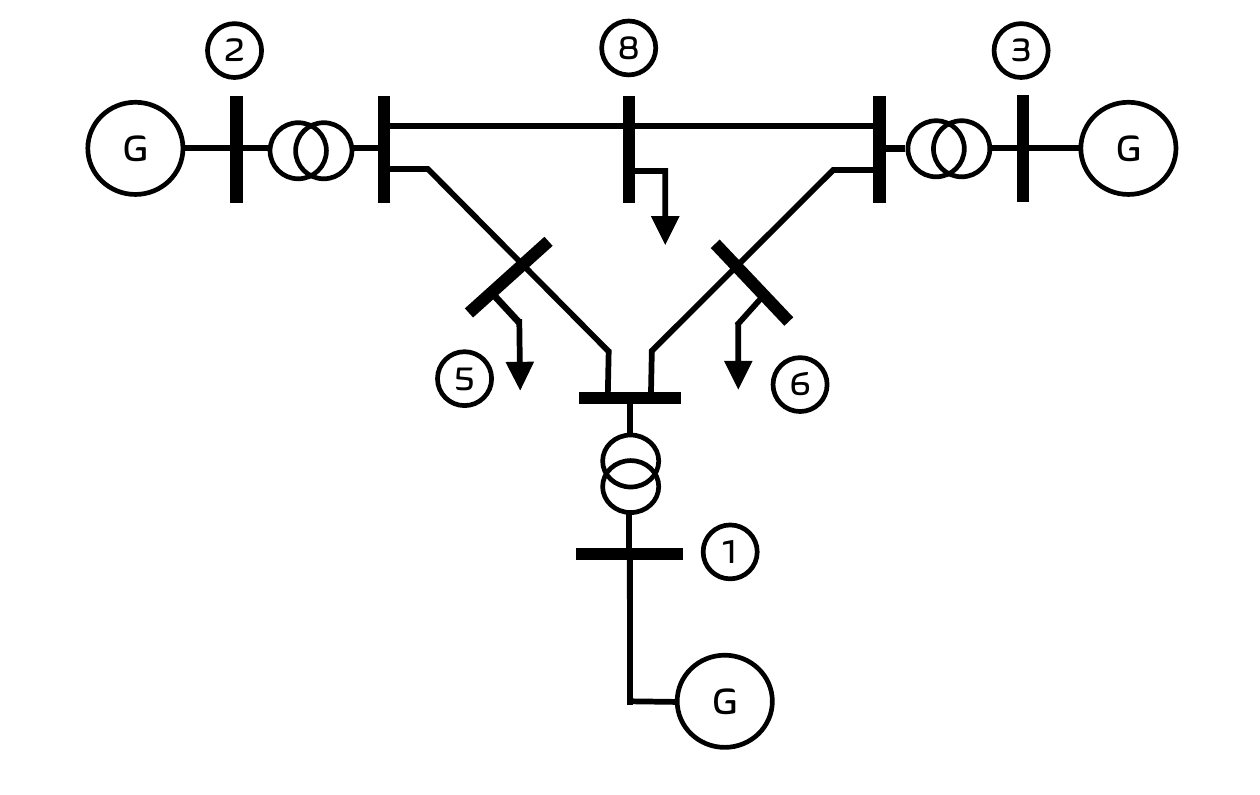}
    \caption{WECC 9-bus system.}
    \label{fig:bus9_system_SG}
\end{figure}

The reduced power flow equation is then: 
   
\begin{equation}\label{eq:dPG}
     \Delta \mathbf{p}_\Gamma  = \mathbf{B}_{r}\Delta\theta_\Gamma + \Delta \mathbf{p}_D,
\end{equation}

 where $\Delta p_{\gamma_i}$, $\Delta \omega_i$, and $\Delta p_{d_i}$ represent the net change in active power injection, change in voltage angle, and change in active power injection resulting from a disturbance at generator node $i$, respectively. The reduced Laplacian, $\mathbf{B}_{r}$, encodes the equivalent susceptances between generator nodes and is calculated from the full susceptance matrix, $\mathbf{B}$, using \eqref{eq:kron}. 
 \begin{equation}\label{eq:kron}
     \mathbf{B}_{r} = \mathbf{B}_{\gamma\gamma}- \mathbf{B}_{\gamma\beta}\mathbf{B}_{\beta\beta}^{-1}\mathbf{B}_{\beta\gamma}
 \end{equation}
The dynamic relationship between an incremental change in active power and an incremental change in voltage angle for a droop GFM is given in \eqref{ode_f_gfm}.
\begin{align}\label{ode_f_gfm}
    \begin{split}
        &\frac{\rmd }{\rmd t} \Delta  \theta_{\gamma_i} = \omega_0 \Delta \omega_i \\[3mm]
        &\frac{\rmd }{\rmd t} \Delta  \omega_i = -\dfrac{1}{T_{ci}}\Delta\omega_i + \dfrac{\alpha_iR_i}{T_{ci}}\Delta p_{\gamma_i}
    \end{split}
\end{align}
Note that the incremental voltage angle and frequency states, $\Delta \theta_{\gamma_i}$ and $\Delta \omega_i$, correspond to generator bus $i$. For $i \in \gamma$ where there are multiple generators at the same bus, \eqref{ode_f_gfm} represents the frequency dynamics of the fictitious single-device equivalent, and parameters with an $i$ subscript are the parameters of this equivalent device. $R_i$ is the droop coefficient, $T_{ci}$ is the frequency response time constant, and $\omega_0$ is the nominal frequency. The parameter $\alpha_i$ is equal to the system base divided by the rated capacity of equivalent device ($\frac{S_B}{S_i}$) For a synchronous generator, this relationship is  \eqref{ode_f_sg}. 
\begin{align}\label{ode_f_sg}
    \begin{split}
        &\frac{\rmd }{\rmd t} \Delta  \theta_{\gamma_i} =  \omega_0 \Delta \omega_i \\[3mm]
        &\frac{\rmd }{\rmd t} \Delta  \omega_i = - \dfrac{D_i}{M_i}\Delta \omega_i - \dfrac{1}{M_i}\Delta p_{m_i} + \dfrac{\alpha_i}{M_i} \Delta p_{\gamma_i} \\[3mm]
        &\frac{\rmd }{\rmd t}\Delta  p_{m_i} = \dfrac{K_i}{T_{SGi}R_{SGi}} \Delta \omega_i -\dfrac{1}{T_{SGi}} \Delta  p_{m_i}
    \end{split}
\end{align}
Here, $M_i$ is the momentum and $D_i$ is the damping coefficient. The governor and turbine response gain constant is $K$, $R_{SGi}$ is the droop constant, and $T_{SGi}$ is the governor and turbine response time constant. The state $\Delta p_{m_i}$ is the incremental power output of the turbine governor. The network model, \eqref{eq:dPG}, is combined with the node-level dynamics, \eqref{ode_f_gfm}-\eqref{ode_f_sg}, to form a linear time-invariant system of the form \eqref{eq:lti}, which is solved exactly using \eqref{eq:soln}.
\begin{equation}\label{eq:lti}
 \dot{x} = \mathbf{A}x+\mathbf{B}u
\end{equation}
\begin{equation}\label{eq:soln}
    x(t) = e^{\bA t}x_0 + \int_{t_0}^t e^{\bA(t - \tau)} B u(\tau) \ \rmd \tau
\end{equation}
Further details on forming \eqref{eq:lti} and \eqref{eq:soln} are given in \cite{Trujillo2025}. Note that in this framework, GFLs are assumed to not have grid support functionality and are modeled as negative loads \cite{Trujillo2025}. Consider the WECC 9-bus system, shown in Figure \ref{fig:bus9_system_SG}. Two common causes of power imbalances are load steps and generator outages. To simulate a load step using the framework in \eqref{eq:dPG}, all load nodes are eliminated, and the effect of the load step is represented as external forcing in the reduced network. For cases where the disturbance node is not in the set of generator nodes, $\mathbf{p}_D$ can be calculated using \eqref{eq:pd}.
\begin{equation}\label{eq:pd}
   \Delta \mathbf{p}_D = \mathbf{B}_{\gamma\beta}\mathbf{B}_{\beta\beta}\Delta \mathbf{p}_\beta,
\end{equation}

\begin{figure}[htpb]
    \centering
    \includegraphics[width=1\linewidth]{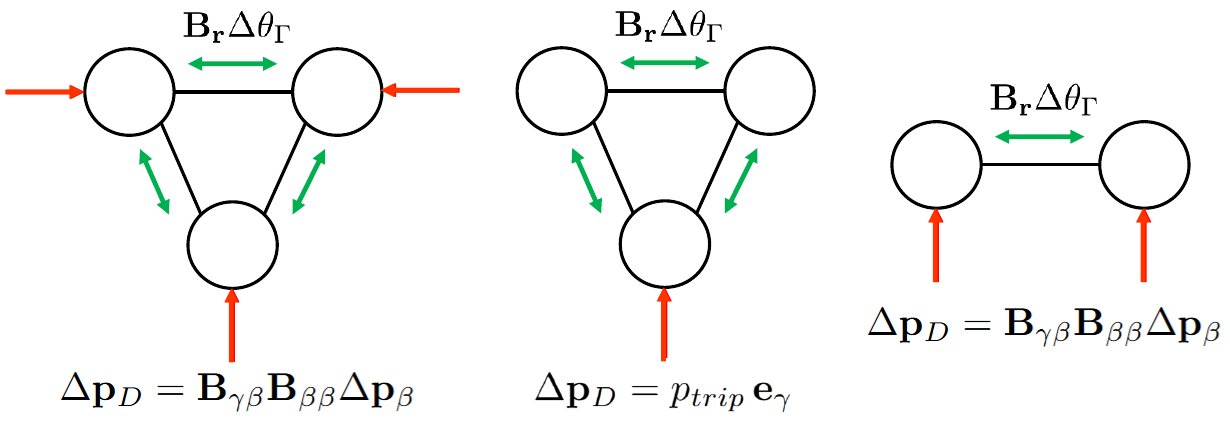}
    \caption{Frequency response model configurations for different disturbance types in the WECC 9-bus system: load step (left), outage at a multi-generator bus (middle), outage at a single-generator bus (right).}
    \label{fig:disturbance_cases}
\end{figure}

In \eqref{eq:pd}, $\Delta \mathbf{p}_\beta$ is the vector of active power disturbance magnitudes occurring at non-generator nodes. Generator outages are represented in a different manner. Suppose there are two generators operating in parallel at one of the generator buses in Figure \ref{fig:bus9_system_SG} and one of the generators is tripped. In this case, the bus where the outage occurs is retained in the network since there is another active generator at the outage bus. The effect of the generator outage is represented as external forcing applied only to the dynamic node where the outage occurred, as shown in the middle image in Figure \ref{fig:disturbance_cases}. Since the disturbance occurs at a generator (retained) node, then the active power disturbance vector, $\Delta \mathbf{p}_D$, is \eqref{eq:ptrip}.

\begin{equation}\label{eq:ptrip}
   \Delta \mathbf{p}_D = p_{trip}\,\mathbf{e}_{\gamma}
\end{equation}

As discussed in Section \ref{sec:apptoUCED}, the largest generator contingency is simulated for each set of hourly commitment and dispatch decisions. Therefore, in \eqref{eq:ptrip}, $p_{trip}$ is equal to the active power injection by the generator with the largest power output at time period $t$ ($p_{trip} = \max_{g \in \mg \cup \ms} p_{g,t}$ for a set of generators, $\mg$, and a set of batteries, $\ms$). Furthermore, $\mathbf{e}_{\gamma} \in \mathbb{R}^{|\gamma|}$ is the standard basis vector selecting the node of the tripped generator. At the bus where the trip occurs, the scaling factor, $\alpha_i$, is modified to reflect that the capacity of the tripped generator is subtracted from the capacity of the single-device dynamic equivalent of that bus.

In contrast, if a generator outage occurs at a node where there is a single generator, that node is eliminated along with other non-generator nodes; it belongs to set $\beta$. In this final scenario, the effect of the outage is represented as if it is a load step via the term $\mathbf{B}_{\gamma\beta}\mathbf{B}_{\beta\beta}\Delta \mathbf{p}_\beta$, as depicted in the right image in Figure \ref{fig:disturbance_cases}. In each case, the first term of \eqref{eq:dPG}, $\mathbf{B_r}\Delta \theta_\Gamma$, represents the power exchanged between generator nodes as their frequencies re-synchronize following a disturbance. In this paper, the largest generator contingency is simulated for each set of hourly commitment and dispatch decisions.

\subsection{Optimization Problem}

The production cost model was created in PowerSimulations.jl \cite{lara2024powersimulations} and follows the mathematical form of Guerra et al. \cite{guerra2025towards}. For a set of thermal generators, $\mg_T$, a set of batteries, $\ms$, over a set of time periods, $\mt$, the objective function is defined as: 
\begin{align}\label{eq:PCM_obj}
    \min &\; \sum_{t \in \mathcal{T}} \sum_{g \in \mathcal{G}_T} c^{fuel}_{g, t} p_{g, t} + c^{start}_{g,t} x^{start}_{g,t} + c^{stop}_{g,t} x^{stop}_{g,t} 
\end{align}
where $p_{g,t}$ is the power from generator $g$ and $c$ are scalar cost values associated with their respective variables. 

For the set of buses, $\mn$, and a set of reserve requirements, $\mr$, (including zonal spinning reserves and system-wide regulation reserves) the constraints on generation are given by: 
\begin{subequations}\label{eq:PCM_gen}
\begin{align}
    &\; \sum_{g \in \mg} p_{g,t} + \sum_{s \in \ms} p^d_{s,t} = \sum_{n \in \mn}d_{n, t} + \sum_{s \in S} p^c_{s,t}, \quad t \in \mt \label{eq:PCM_system_balance} \\
    &\; \underline{P}_{g,t} x_{g,t} \le p_{g,t} + \sum_{i \in \mr} r_{i, g, t} \le \overline{P}_{g,t} x_{g,t}, \quad g \in \mg, t \in \mt \label{eq:PCM_gen_limits} \\
    &\; \underline{R}_{g,t} \le p_{g, t} - p_{g, t-1} \le \overline{R}_{g,t}, \quad g \in \mg_T, t \in \mt \label{eq:PCM_ramp_rates} \\
    &\; x_{g, t-1} - x_{g, t} + x^{start}_{g,t} - x^{stop}_{g, t}  = 0, \quad g \in \mg_T, t \in \mt \label{eq:PCM_gen_binaries} \\
    &\; \sum_{g \in \mg} r_{i, g, t} + \sum_{s \in \ms} r_{i,s,t} \ge RR_{i, t}, \quad i \in \mr, t \in \mt \label{eq:PCM_reserve_requirements} \\
    &\; \underline{Q}_{i,g,t} \le r_{i,g,t} \le \overline{Q}_{i,g,t} \quad g \in \mg \cup \ms, i \in \mr, t \in \mt \label{eq:PCM_reserve_limits}
\end{align}
\end{subequations}
where $\mathcal{G}$ is the set of all generators (thermal and renewable), $d_{n,t}$ are the demands at bus $n$, $\underline{\cdot}$ and $\overline{\cdot}$ are lower and upper bounds on their respective variables or expressions, $p^d_{s,t}$ is the power discharged from battery $s$, $p^c_{s,t}$ is the power used to charge battery $s$, $x_{g,t}$ is a binary variable indicating if generator $g$ is dispatched, $x^{start}_{g,t}$ and $x^{stop}_{g,t}$ are binary variables indicating if generator $g$ started up or shut down in the time period $t$, $r$ is the amount of power contributed to the reserve requirement, and $RR_{i,t}$ is the reserve requirement of reserve category $i$. Here, \eqref{eq:PCM_system_balance} is the system power balance, \eqref{eq:PCM_gen_limits} is the generator power production limits, \eqref{eq:PCM_ramp_rates} are the limits on ramp rates, \eqref{eq:PCM_gen_binaries} constrain the start up and shut down of the generators, \eqref{eq:PCM_reserve_requirements} set the reserve requirements, and \eqref{eq:PCM_reserve_limits} are limits on the reserve variables. 

Line flow constraints in the system are given by: 

\begin{subequations}\label{eq:PCM_lines}
\begin{align}
    &\; \begin{aligned}[t]
        \sum_{(j,k) \in \ml | j = n} f_{j,k,t} - \sum_{(j,k) \in \ml | k = n} f_{j,k,t} = \sum_{g \in \mg(n)} p_{g,t} - d_{n, t} + \\ \sum_{s \in \ms(n)} p^d_{s,t} -  \sum_{s \in \ms(n)} p^c_{s,t}, \quad n \in \mn, t \in \mt 
    \end{aligned} \label{eq:PCM_nodal_balance} \\
    &\; \underline{F}_{i,j} \le f_{i,j,t} \le \overline{F}_{i,j}, \quad (i,j) \in \ml, t \in \mt \label{eq:PCM_flow_bounds} \\
    &\; f_{j,k,t} = B_{j,k} (\theta_{j,t} - \theta_{k,t}), \quad (i,j) \in \ml, t \in \mt \label{eq:PCM_DCOPF_flows} \\ 
    &\; \underline{\theta} \le \theta_{n, t} \le \overline{\theta}, \quad n \in \mn, t \in \mt \label{eq:PCM_angle_bounds} \\
\end{align}
\end{subequations}
where $\mg(n)$ and $\ms(n)$ are the generators and batteries on bus $n$, respectively; $\ml$ is the set of lines, $f_{j,k,t}$ is the flow between buses $j$ and $k$, $B_{j,k}$ is the susceptance of the respective line, and $\theta_j$ is the bus phase angle on bus $j$. Here, \eqref{eq:PCM_nodal_balance} is the nodal power balance, \eqref{eq:PCM_flow_bounds} are the bounds on flows, \eqref{eq:PCM_DCOPF_flows} are DCOPF constraints on power flows, and \eqref{eq:PCM_angle_bounds} are bounds on the phase angles. 

Finally, the battery storage operation constraints are given by:
\begin{subequations}\label{eq:PCM_storage}
\begin{align}
    &\; 0 \le p^c_{s,t} \le \overline{P}^c_{s} x^c_{s,t}, \quad s \in \ms, t \in \mt \label{eq:PCM_charge_bound}\\
    &\; 0 \le p^d_{s,t} \le \overline{P}^d_{s} (1 - x^c_{s,t}), s \in \ms, t \in \mt \label{eq:PCM_discharge_bound} \\
    &\; SoC_{s,t} = (1 - \eta^{sd}_s) SoC_{s,t-1} + \frac{p^{c}_{s,t}}{\eta^c_{s}} - \eta^d_{s} p^d_{s,t}, \quad s \in \ms, t \in \mt \label{eq:PCM_SOC_by_time}\\
    &\; \underline{SoC}_s \le  SoC_{s, t} \le \overline{SoC}_s, \quad s \in \ms, t \in \mt \label{eq:PCM_SOC_bounds}\\
    &\; SoC_{s,t=1} = SoC^{init}_{s}, \quad s \in \ms \label{eq:PCM_SOC_init} \\
    &\; p^d_{s,t} + \sum_{i \in \mr} r_{i,s,t} \le \overline{P}^d_{s} + p^c_{s,t}, \quad s \in \ms, t \in \mt \label{eq:PCM_storage_reserves}
\end{align}
\end{subequations}

where $x^c_{s,t}$ is an auxiliary variable to ensure that the battery cannot charge and discharge at the same time, $SoC_{s,t}$ is the state of charge of battery $s$, $SOC^{init}_s$ is a parameter of the initial state of charge of the battery, and $\eta^{sd}_{s}$, $\eta^{c}_s$, and $\eta^d_{s}$ are the self discharge rate, charging, and discharging efficiencies for battery $s$, respectively. Here \eqref{eq:PCM_charge_bound} and \eqref{eq:PCM_discharge_bound} are bounds on the charging and discharging of the battery, \eqref{eq:PCM_SOC_by_time} is the temporal operation of the battery, \eqref{eq:PCM_SOC_bounds} are bounds on the state of charge, \eqref{eq:PCM_SOC_init} sets the initial state of charge of the battery, and \eqref{eq:PCM_storage_reserves} defines the battery's contribution to reserves.

In addition to the noted constraints, there are also slack variables on the system balance and the reserve requirements, which are heavily penalized in the objective function. All variables are nonnegative. Finally, we note that, for the use of the branch and cut algorithm, we added a binary dispatch variable for short-duration storage devices when discharging.

Within PowerSimulations.jl, the optimization problem is run for a day-ahead period of 24 hours with hourly resolution and an additional 24 hours of ``look-ahead'' horizon ($|\mt| = 48$). Simulations are run one day at a time by solving the optimization problem, keeping the first 24 time points, then moving forward in time and continuously maintaining a 48-hour time horizon. The results of the previous day are also fixed at the start of the next days simulation, such that the commitment status of the generators must be maintained when solving for the next day. Each step of the simulation was solved with Gurobi using lazy constraint callbacks which are called at each new incumbent solution. At each new incumbent, the frequency response model with the largest generator contingency was evaluated (see Section III), and if it was infeasible at any hour, a constraint was added to that hour to cut off that incumbent solution and add an additional constraint on i) thermal generation or ii) the power output of a critical generator, to push the model towards feasible solutions. Because the second constraint is applied to continuous variables, it was possible that an incumbent solution could become feasible after the constraint was added. Consequently, we added binary variables to the system that do not show up in the objective or operational constraints, but instead allow the solver to find the same operational incumbent solution again. After each step of the simulation, the lazy constraints were removed from Gurobi so that the next step began with no lazy constraints. 
\section{Case Study}\label{casestudy}

Here we present results from the application of the problem to a 73-bus system considering different sets of constraints. 
\subsection{Results - Base Case}\label{base case}
\begin{figure}[htpb]
    \centering
    \includegraphics[width=.7\linewidth]{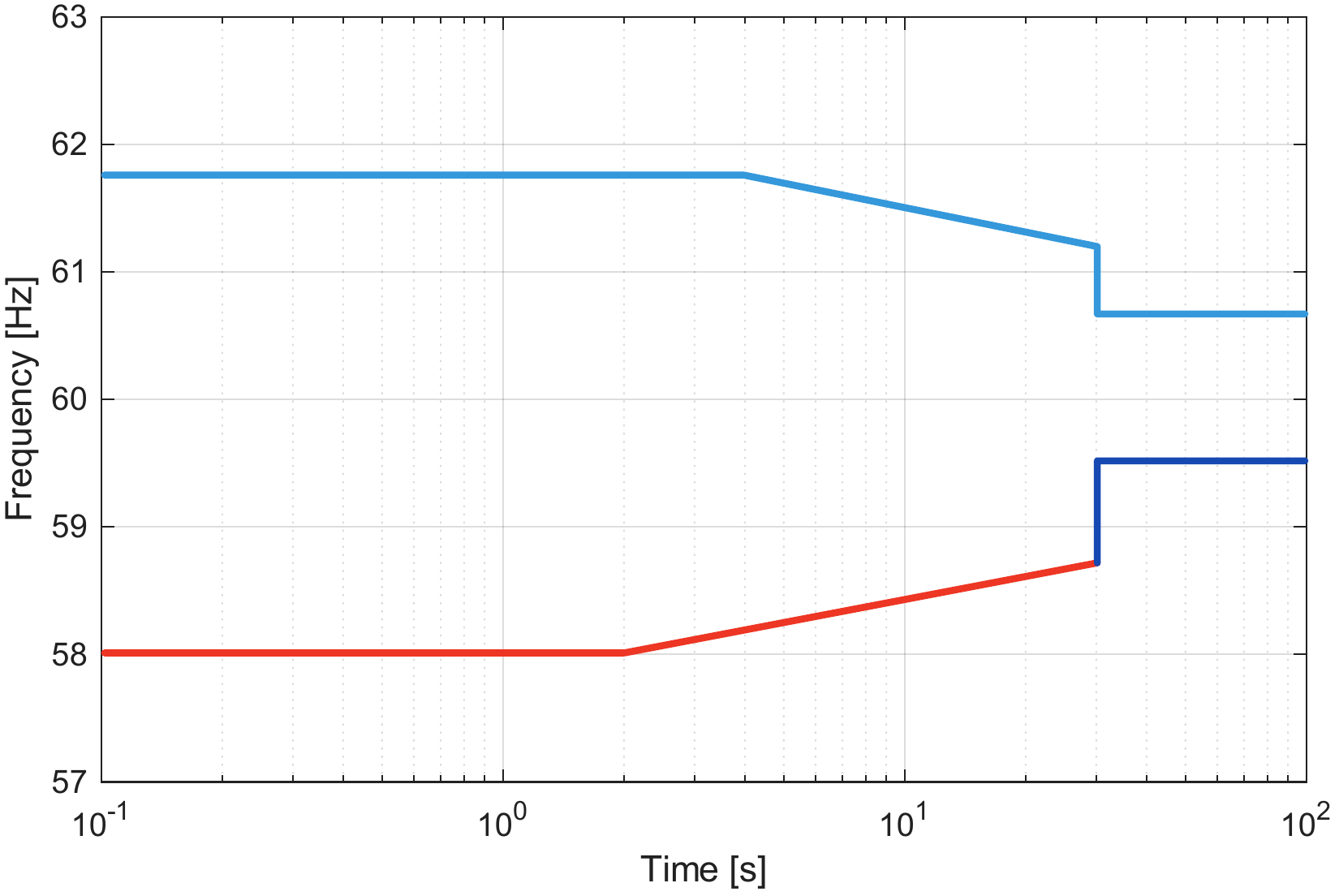}
    \caption{Frequency bounds for the Eastern Interconnection \cite{PRC006NPCC2AutomaticUnderfrequency2019}. The continent-wide underfrequency requirement for $<30$s outlined in NERC PRC-006 is shown in red.}
    \label{fig:ufls_limits}
\end{figure}

In this section we present the results of the optimization problem applied to a modified version of the RTS-GMLC system \cite{barrowsIEEEReliabilityTest2020, RTS_data}, which is a 73-bus system. \textcolor{black}{The installed capacity values by generation type are provided in Table \ref{tab:capacity}. We henceforth assume that every IBR is of the droop GFM type and is referred to as a ``GFM" for simplicity} The system was modified such that at every generation bus, if the total capacity of SG assets exceeded that of the GFMs, the GFMs were removed and their capacity was added to the SG assets. Conversely, if the total capacity of the GFMs exceeded that of the SGs, the SGs were removed and their capacity was added to the GFMs. These modifications were necessary for the application of the frequency model proposed in \cite{Trujillo2025} to the system, since the frequency model assumes each generation bus is either an ``GFM bus" or a ``SG bus." In addition, a short-duration storage unit was assumed to be an GFM and was moved to Bus $316$ so that there were no mixed-generation-type buses in the system, and the long-duration storage element was disabled.  \textcolor{black}{It is well known that GFMs must operate with sufficient headroom to participate in frequency response \cite{linResearchRoadmapGridForming}, and therefore it is assumed that generation assets are operating with sufficient headroom as to be able to participate in frequency response.} All GFMs are assumed to have $P$-$\delta$ droop characteristics as indicated in \eqref{ode_f_gfm}; dead bands are not modeled. All GFMs are assumed to have identical droop coefficients and time constants while SGs have identical inertia, damping, and droop coefficients, as well as turbine governor time constants. We note that GFM and SG parameters are not required to be identical in the frequency model proposed in \cite{Trujillo2025}, and that identical parameters are assumed in this paper in the interest of simplicity. The rated capacities of the generation assets are not identical. The instantaneous frequency threshold was chosen to be $58$ Hz, in accordance with the continent-wide underfrequency requirement laid out by NERC PRC-006 and depicted in Figure \ref{fig:ufls_limits}. Finally, in this framework, we focus on low instantaneous frequencies; low settling frequencies that would be managed by automatic generation control are treated as a separate problem for future work, though reserves are held, but not deployed, for this purpose in the model.

\begin{table}
    \centering
    \begin{tabular}{c|c|c}
       Generation Type & Installed Capacity [MVA] &Grid Interface\\ \hline
       Gas, Steam & 8681&SG\\
       Nuclear & 447&SG\\
       Solar PV & 8611&GFM\\
       Wind &3307&GFM\\
       Short-Term Battery & 1000&GFM\\
    \end{tabular}
    \caption{Installed capacity by generation type. Grid interface is also indicated.}
    \label{tab:capacity}
\end{table}
\begin{figure}[htpb]
    \centering
    \includegraphics[width=.4\linewidth]{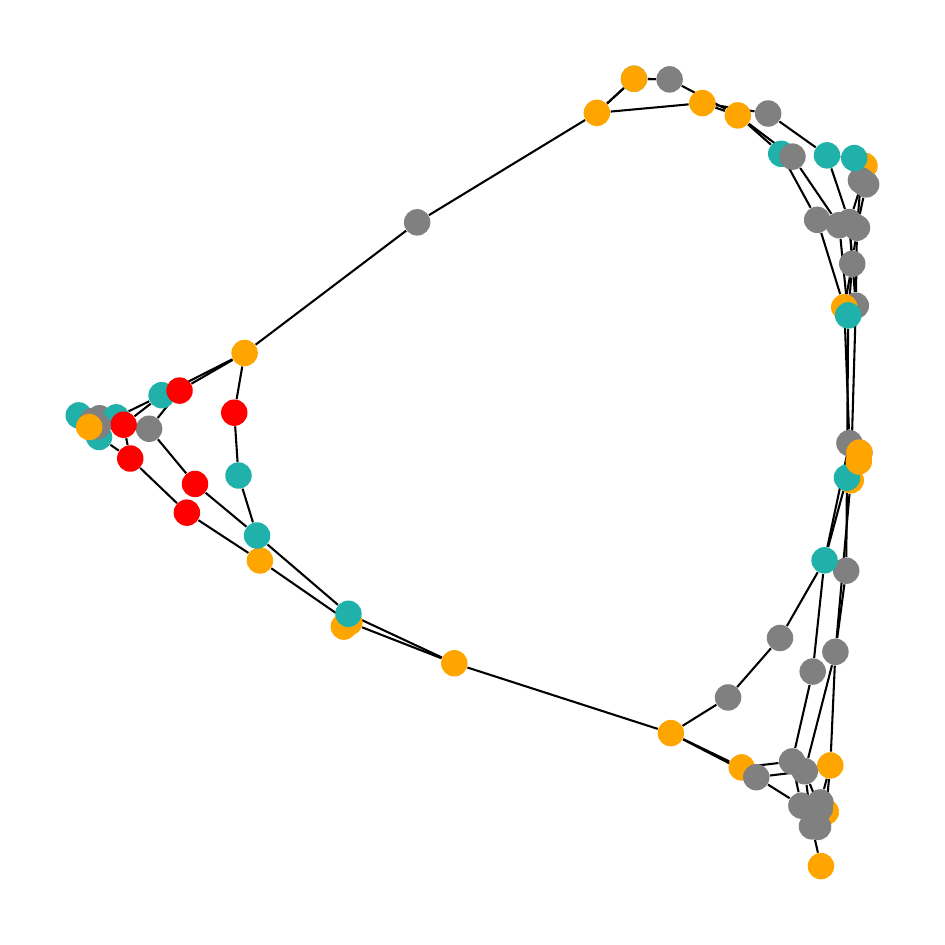}
    \caption{Graph representation of the modified RTS-GMLC system. GFM-buses are shown as blue nodes and SG-buses as orange nodes. Nodes where the minimum frequency went below $58$Hz in the full-year problem are shown in red. All of the frequency-unstable nodes were GFM nodes. Gray nodes represent buses without generation devices.}
    \label{fig:rts_graph}
\end{figure}

\begin{figure}[htpb]
    \centering
    \begin{subfigure}[c]{0.5\textwidth}
        \centering
        \includegraphics[width=0.85\linewidth]{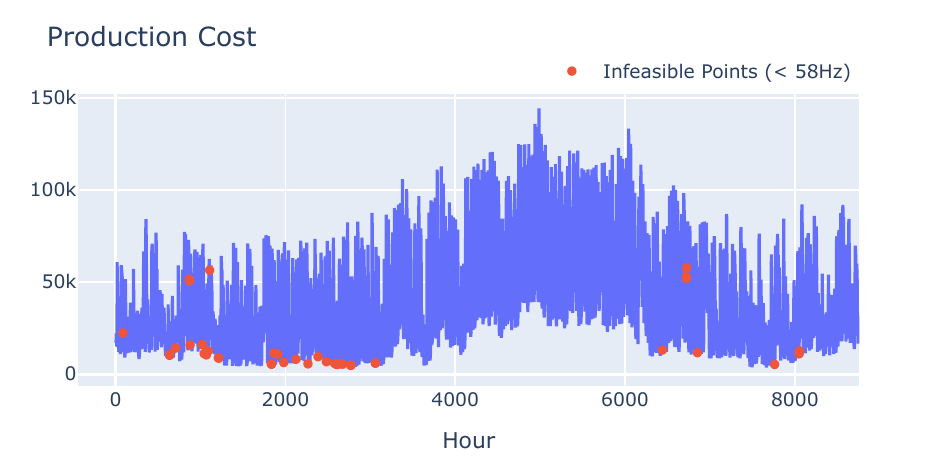}
        \caption{Production cost for the unconstrained case over a full-year. Time periods when the minimum frequency in the system was below $58$ Hz are shown in red.}
        \label{fig:p_cost_58}
    \end{subfigure}
    \begin{subfigure}[c]{0.5\textwidth}
        \centering
        \includegraphics[width=0.85\linewidth]{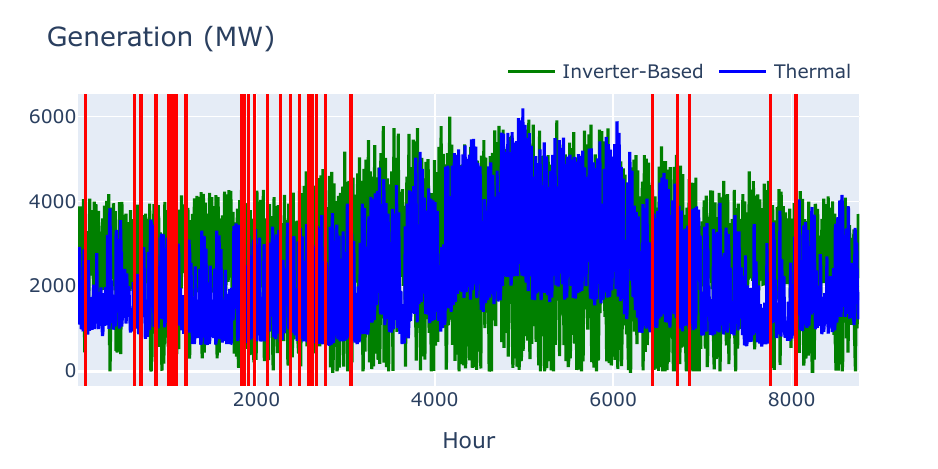}
        \caption{Dispatch levels of thermal and inverter-based resources over a full-year for the base case (without frequency constraints). Time periods when the minimum frequency in the system was below $58$ Hz are shown in red.}
        \label{fig:p_dispatch_58}
    \end{subfigure}
    \caption{Results for one year. Instances of excessively low frequency values generally correspond to time periods with high GFM to SG power output ratios. }
    \label{fig:basecase}
\end{figure}

We first present the results of a full-year UC \& ED problem without frequency constraints. The production cost over the full year is shown in Figure \ref{fig:p_cost_58}. Time periods when the minimum frequency in the system would have been below the instantaneous UFLS level ($58$Hz) in the event of a largest generator contingency are shown in red; there were $43$ of these time periods over the course of the year. Importantly, these infeasible points occur during times of high instantaneous GFM shares, and thus relatively low SG shares, as reflected in the dispatch curves depicted in Figure \ref{fig:p_dispatch_58}.

\begin{figure}[htpb]
	\centering
	\begin{subfigure}[b]{0.5\textwidth}
		\centering\includegraphics[width=0.85\textwidth]{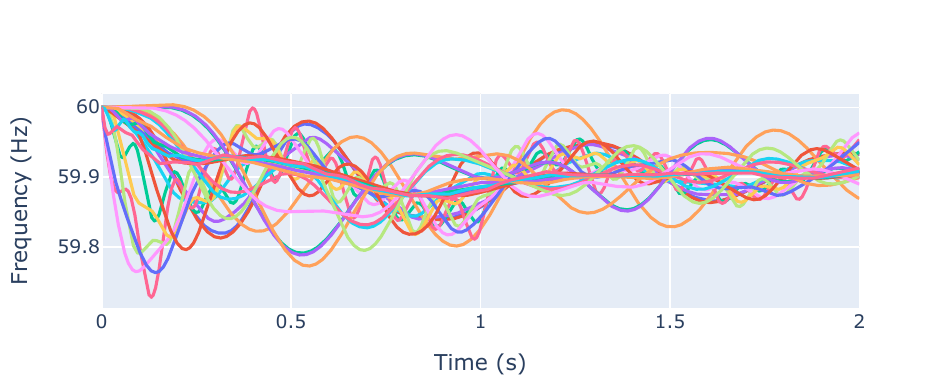}
		\caption{Example of a dispatch scenario where all frequencies stayed above the instantaneous UFLS limit (Hour 5000).}
		\label{fig:normalfreq_example}
	\end{subfigure}
	\hspace{10pt}
	\begin{subfigure}[b]{0.5\textwidth}
		\centering\includegraphics[width=0.85\textwidth]{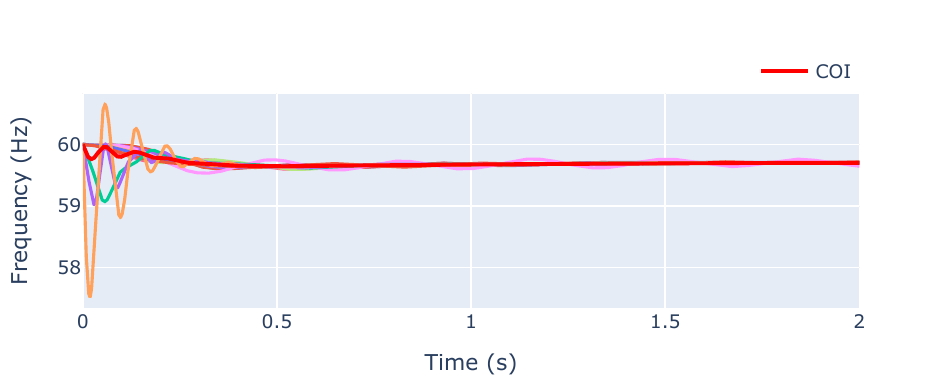}
		\caption{Example of a dispatch scenario where the instantaneous UFLS limit was violated (Hour 636). The red trace represents the average (COI) frequency.}
		\label{fig:lowfreq_example}
	\end{subfigure}
	
	\caption{Frequency response at every generation node after the largest generator contingency for two dispatch scenarios.}
	\label{fig:freqs_base}
\end{figure}

The frequency response plots at two different time instances are shown in Figure \ref{fig:freqs_base}. These plots show the frequency trajectories at every online generator bus in the system. In Figure \ref{fig:normalfreq_example}, which corresponds to hour 5000 in the full-year UC \& ED results, no buses violated the instantaneous frequency  limit of $58$ Hz. In Figure \ref{fig:lowfreq_example}, however, which corresponds to hour 636, one of the nodal frequencies does violate the frequency limit. Notably, the average (COI) frequency, which is shown in red, remains close to the nominal frequency. The severe overestimation of the minimum instantaneous nodal frequency by the COI frequency metric is a fundamental limitation of approximating frequency response by a single trajectory. In fact, for every infeasible point in the full-year simulation, the COI frequency did not predict an instantaneous frequency below the underfrequency performance characteristic curve. The COI frequency in Figure \ref{fig:lowfreq_example} was directly calculated by taking the average of all the frequency trajectories calculated by the frequency model introduced in Section \ref{overview}. In practice, analytically-derived system frequency models have difficulty in capturing the rich dynamics of mixed-generation systems and neglect the impact of disturbance location, adding additional sources of error \cite{chan2007dynamic,Anderson1990ALS,shi2018analytical,liuAnAnalyticalModel}.


In Sections \ref{app1} and \ref{app2}, we present two methods of constraining the UC\&ED problem such that instantaneous nodal frequencies do not fall below $58$ Hz. These constraints are heuristic and do not guarantee the lowest possible operating cost which does not violate stability requirements, but they provide solutions that are feasible while maintaining stability. The first approach aims to mitigate low frequencies  by increasing the thermal generation in the system, thereby increasing inertia. The second approach decreases the power output of a critical generator. Results of the sensitivity analysis are presented in Section \ref{sensitivity} and a discussion on the efficacy of the different approaches is discussed in Section \ref{discussion}.


\subsection{Approach 1 - Constraint on Thermal Generation}\label{app1}

The results of the unconstrained case suggest that a lack of online SGs is a contributing factor to frequency trajectories that violate the instantaneous frequency limit. For this reason, we introduce a constraint on the minimum thermal generation for incumbent solutions that violate the frequency stability requirements. The constraint is given in \eqref{thermal_constraint} for a time point $t$ where the frequency stability requirements are violated and where $^*$ corresponds to the value at the incumbent solution. 

\begin{equation}\label{thermal_constraint}
    \sum_{g \in \mg_T} p_{g,t} \geq 1.05\sum_{g \in \mg_T} p^*_{g,t}
\end{equation}

\begin{figure}[h]
    \centering
    \begin{subfigure}[c]{0.5\textwidth}
        \centering
        \includegraphics[width=0.85\linewidth]{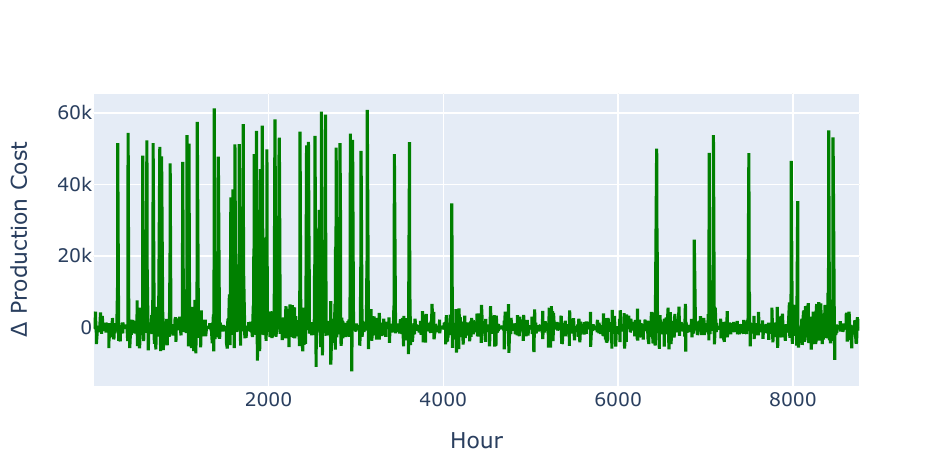}
        \caption{Difference in production cost over one year (thermal generation constrained production cost $-$ base case production cost).}
        \label{fig:p_diff_thermal}
    \end{subfigure}
    \begin{subfigure}[c]{0.5\textwidth}
        \centering
        \includegraphics[width=0.85\linewidth]{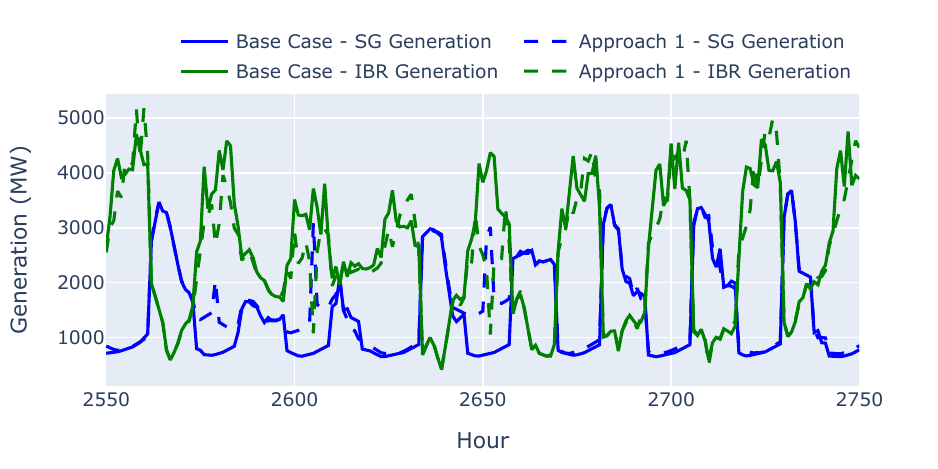}
        \caption{Comparison of hourly generation from GFMs and SGs in the unconstrained case and the thermal generation-constrained case.}
        \label{fig:gen_app1_base}
    \end{subfigure}
      \begin{subfigure}[c]{0.5\textwidth}
        \centering
        \includegraphics[width=0.85\linewidth]{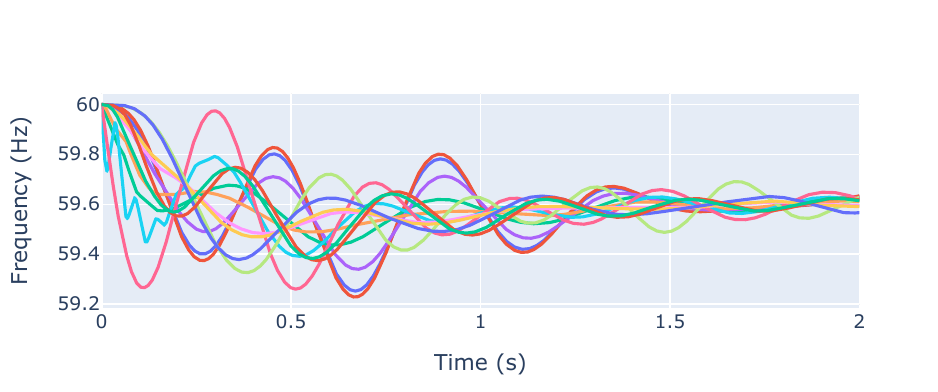}
        \caption{Frequency response at every generation node for hour 636 (infeasible in the base case).}
        \label{fig:lowfreq_example_t}
    \end{subfigure}
    \caption{Results for one year. Infeasible solutions were cut off and constraints on thermal generation were implemented. }
    \label{fig:thermal_results}
\end{figure}

The results of the thermal generation constrained problem are shown in Figure \ref{fig:thermal_results}. The new solutions are free from unstable frequency scenarios as predicted by the frequency model. Comparison of the production costs between the unconstrained case and the thermal generation-constrained case reveals that production costs soar during the ``infeasible" hours indicated in Figure \ref{fig:basecase}. The difference in cost (where the production cost in the base case was subtracted from the cost in the thermal generation-constrained case) is given in Figure \ref{fig:p_diff_thermal}. The increased costs are due the shifting of generation from zero-marginal cost GFMs to more expensive SG-based devices. The hourly generation from thermal/SG devices and GFMs for both the base case and the constrained case is illustrated in Figure \ref{fig:gen_app1_base}. The frequency trajectories for Hour 636, which was an infeasible time period in the base case, are shown in Figure \ref{fig:lowfreq_example_t}. Qualitatively, the frequency curves are far different from those seen in the base case (Figure \ref{fig:lowfreq_example}).

\subsection{Approach 2 - Critical Generator Output Constraint}\label{app2}
\begin{figure}[htpb]
    \centering
    \begin{subfigure}[c]{0.5\textwidth}
        \centering
        \includegraphics[width=0.9\linewidth]{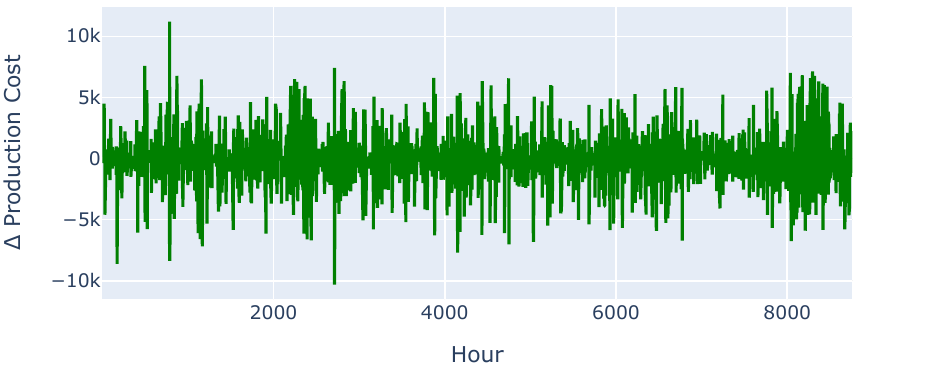}
        \caption{Difference in production cost over one year (frequency constrained production cost $-$ base case production cost).}
        \label{fig:p_diff}
    \end{subfigure}
    \begin{subfigure}[c]{0.5\textwidth}
        \centering
        \includegraphics[width=0.9\linewidth]{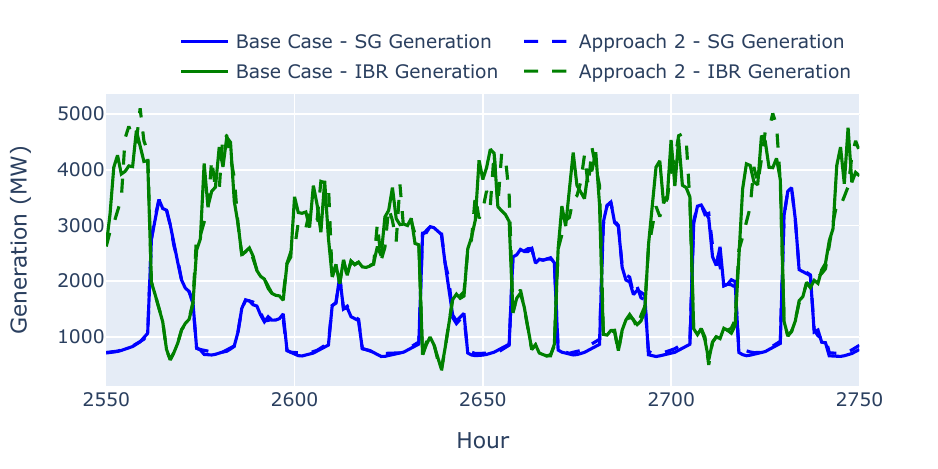}
        \caption{Comparison of hourly generation from GFMs and SGs in the unconstrained case and the critical generator output-constrained case.}
        \label{fig:gen_app2_base}
    \end{subfigure}
      \begin{subfigure}[c]{0.5\textwidth}
        \centering
        \includegraphics[width=0.9\linewidth]{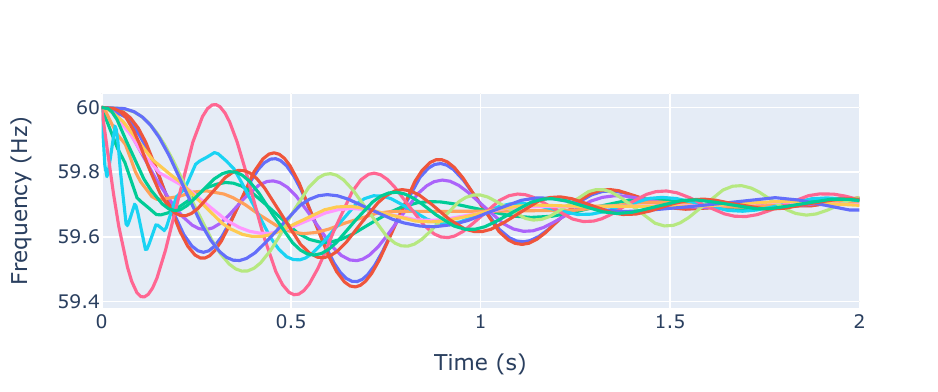}
        \caption{Frequency response at every generation node for hour 636 (infeasible in the base case). }
        \label{fig:lowfreq_example_c}
    \end{subfigure}
    \caption{Results for one year. Infeasible solutions were cut off and constraints on the critical generator output were implemented. }
    \label{fig:main}
\end{figure}

The elevated production costs caused by increased thermal generation motivate the formulation of a second approach, which seeks to encourage frequency-stable solutions which have a diminished impact on production cost. In this second approach, a constraint is placed on the maximum power output of the ``critical generator," which is the generator with the largest power output in the incumbent solution. The contingency under study is always the outage of the generator with the largest power output at the specified hour, and thus the constraint encourages a slight redistribution of power output, lessening the impact of the largest generator contingency. The constraint at time $t$ where the frequency stability requirements are violated is give in \eqref{contingency_constraint}.
\begin{equation}\label{contingency_constraint}
     p_{\hat{g},t} \leq 0.9 \cdot p^*_{\hat{g},t}
\end{equation}
\noindent where $\hat{g}$ is the critical generator at time $t$. We then solve the full-year problem to confirm the effectiveness of Approach 2 in eliminating frequencies below $58$ Hz while not increasing production cost significantly. The resulting production costs are compared to those of the base case in Figure \ref{fig:p_diff}.

The hourly generation from GFMs and SGs in the base case and the case using constraint Approach 2 are compared in Figure \ref{fig:gen_app2_base}. In contrast to the solutions obtained using Approach 1, thermal generation is nearly the same as in the base case, while generation from GFMs is slightly shifted in time. The frequency trajectories at hour 636 are given in Figure \ref{fig:lowfreq_example_c}. From these results we see that Approaches 1 and 2 yielded similar, though not identical, solutions for this time period.

The results suggest that the constraint on the  critical generator output outperforms the constraint on thermal generation in several ways. First, the constraint on the critical generator output was effective at mitigating excessively low frequencies without substantially increasing the average daily production cost. Second, the effectiveness of Approach 2 demonstrates that it is possible to reduce low frequencies without bringing more carbon-intensive generation online. These results stand in stark contrast to the results obtained using Approach 1, where dramatic increases in production cost and thermal generation are observed. Further discussion on the effectiveness of each approach is included in Section \ref{discussion}.

\subsection{Sensitivity Analysis}\label{sensitivity}

\begin{figure}[htpb]
    \centering
    \includegraphics[width=0.85\linewidth]{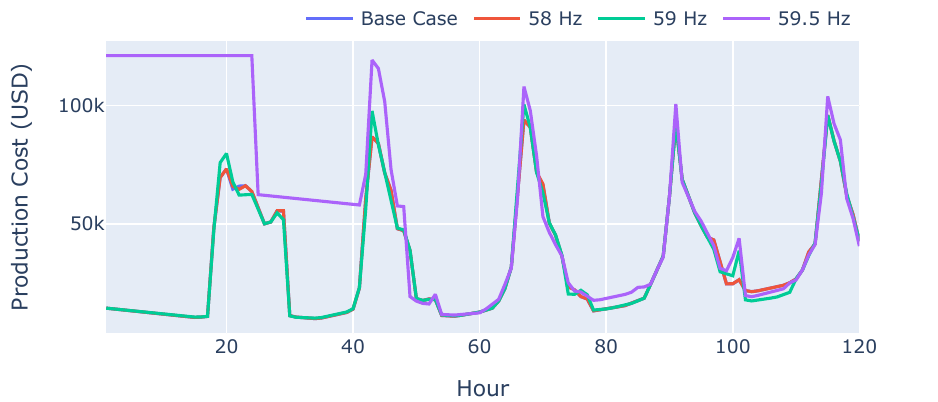}
    \caption{Production cost over five days for three different instantaneous frequency limits.}
    \label{fig:p_cases}
\end{figure}

Next, the sensitivity of the production cost to changes in the instantaneous frequency threshold was evaluated. For this analysis, Approach 2 was used, and the frequency thresholds were $58$, $59$, and $59.5$ Hz. The production cost over five days for these scenarios and the unconstrained case are provided in Figure \ref{fig:p_cases}. The production cost for the unconstrained case and for the frequency-constrained cases using $58$ and $59$ Hz limits were largely the same. However, when the minimum frequency limit was set to $59.5$ Hz there is a sustained increase in production cost in the first two days. This increase in production cost is expected given the tightened constraint on instantaneous frequency and the subsequent increase in infeasible time periods. 

\subsection{Nodal RoCoF Constraints}\label{sec:rocof}

\begin{figure}[htpb] 
    \centering
    \begin{subfigure}[c]{0.5\textwidth}
    \centering\includegraphics[width=0.9\linewidth]{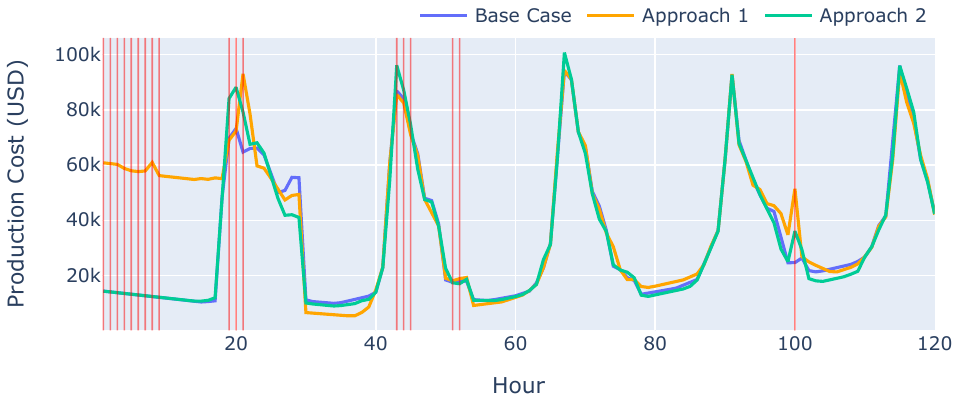}
    \caption{Production cost over five days for the RoCoF-constrained problem with an additional constraint on: 1) thermal generation, and 2) critical generator output. The production cost for the base case is also included for comparison. The red lines indicate time periods when the maximum RoCoF in the system exceeded $1$ Hz/s in the base case.}
    \label{fig:rocof}
    \end{subfigure}
    \begin{subfigure}[c]{0.5\textwidth}\centering
    \includegraphics[width=0.9\linewidth]{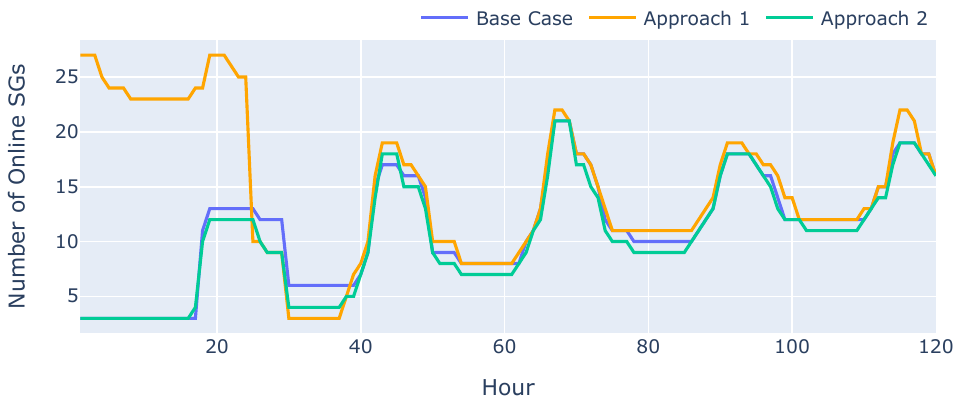}
    \caption{Number of SGs online over five days. The thermal generation-constrained case results in much greater numbers of online SGs compared to the base case and the critical generator output-constrained case.}
    \label{fig:num_SGs}
    \end{subfigure}
    \caption{Five-day results of the RoCoF-constrained problems.}
\end{figure}

Given that unexpected IBR tripping can occur due to fast RoCoFs \cite{MeetingChallengesDeclining}, it is desirable to assess the robustness of the proposed problem formulation for limiting RoCoF. In this scenario, feasibility is determined by calculating local RoCoFs rather than instantaneous frequencies, and cutting planes are generated accordingly. A $0.5$ s window is used for calculating RoCoF. The results of the RoCoF-constrained problem with an additional constraint on: 1) thermal generation, and 2) the critical generator output, are compared to the base case results in Figure \ref{fig:rocof}. The RoCoF threshold was $1$ Hz/s. There were 18 time periods when the maximum RoCoF exceeded $1$ Hz/s in the base case; these points are indicated by red vertical lines.

As was the case in the nodal instantaneous frequency-constrained problem, the critical generator output constraint was more effective than the thermal generation constraint in terms of keeping production cost low while mitigating undesirable frequency behavior. In the thermal generation-constrained case, many more SGs had to be dispatched to provide enough inertia to mitigate high RoCoFs as compared to the base case and the critical generator output-constrained case, as seen in Figure \ref{fig:num_SGs}. Interestingly, using Approach 2 resulted in less SG deployment than the base case, but was still effective in eliminating RoCoFs $> 1$ Hz/s. 

\begin{figure}[htpb]
    \centering
    \includegraphics[width=0.9\linewidth]{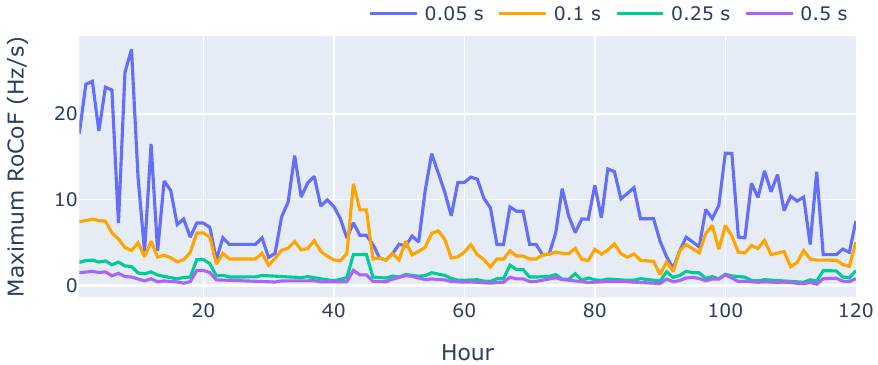}
    \caption{Maximum RoCoF over five days using four different time measurement windows.}
    \label{fig:p_rocof_windows}
\end{figure}

It must be noted that the RoCoF values varied dramatically depending on the time window used for measurement. The maximum RoCoF for each hour using four different time measurement windows is shown in Figure \ref{fig:p_rocof_windows}. Generally speaking, as the time measurement window is shortened, the better the rapid frequency changes at GFM nodes are captured. The filtering effect longer time windows have on fast dynamics  may or may not be desirable depending on the  phenomena of interest. 
%

%
%
\subsection{Discussion}\label{discussion}

In subsections \ref{base case} $-$ \ref{sec:rocof}, the results corresponding to the following problem formulations were presented:
\begin{enumerate}[i.]
    \item Base case (no frequency stability constraints).
    \item Instantaneous nodal frequency constrained-case with an additional constraint on thermal generation.
    \item Instantaneous nodal frequency constrained-case with an additional constraint on the critical generator output.
    \item Nodal RoCoF constrained-case with an additional constraint on thermal generation.
    \item Nodal RoCoF constrained-case with an additional constraint on the critical generator output.
\end{enumerate}

The critical generator output constraint outperformed the thermal generation constraint in terms of keeping production costs low. The solve time of the critical generator output-constrained problem was also considerably faster than the the thermal-generation constraint approach. In Table \ref{tab:avg_dispatch}, the average hourly SG and GFM dispatch, the average daily production cost, and the solve time for problem formulations i-iii, are given. All solve times were obtained on an AMD Ryzen 7 4700U CPU @ 2.00GHz, 8-core processor. Seeing that a standard laptop was used in this study, there are very substantial time savings to be gained by utilizing a high performance computer.
\setlength{\tabcolsep}{2pt}
\begin{table}[h]
    \centering
    \begin{tabular}{c|c|c|c|c}
     &Avg. Hourly SG&Avg. Hourly GFM& Avg. Daily Prod.& Solve\\
       Case &  Dispatch [GW] &  Dispatch [GW]&Cost [\$1M]& Time [hr]\\ \hline
        i & 2.12& 2.37&0.92 &3.16\\
       ii &2.17 &2.31 &0.95 &17.28\\
       iii &2.12 & 2.37& 0.92&11.97\\
    \end{tabular}
    \caption{Average dispatch by grid interface type, average daily production cost, and solve time for each solution set for one full year. The problem formulations are referred to by their place in the enumerated list at the beginning of this subsection.}
    \label{tab:avg_dispatch}
\end{table}

The results of the five-day simulations using formulations iv and v arrived at similar conclusions. That is, when eliminating high RoCoFs instead of low instantaneous frequencies, the constraint on the critical generator output outperformed the constraint on thermal generation. The solutions for formulations i, iv, and v are included in Table \ref{tab:avg_dispatch2}. \textcolor{black}{Note that the marginally lower cost of the formulation in Case \textit{v} over Case \textit{i} is a result of the simulations taking different pathways over the course of sequential simulations}. To concurrently constrain instantaneous frequencies and RoCoFs, a simple AND operator may be employed when determining the feasibility of the solution.
\begin{table}[h]
    \centering
    \begin{tabular}{c|c|c|c|c}
     &Avg. Hourly SG&Avg. Hourly GFM& Avg. Daily Prod.& Solve\\
       Case &  Dispatch [GW] &  Dispatch [GW]&Cost [\$1M]& Time [min]\\ \hline
        i &1.92 & 2.70& 0.81& 6.54\\
       iv & 2.18& 2.44& 0.97& 74.68\\
        v & 1.90& 2.72& 0.80& 23.50\\
    \end{tabular}
    \caption{Results of the five-day UC \& ED. Problem formulations are referred to by their place in the enumerated list at the beginning of this subsection.}
    \label{tab:avg_dispatch2}
\end{table}
\section{Conclusions and Future Work}\label{conclusion}
As the proliferation of renewable technologies continues, frequency stability concerns arise due to the fast dynamics of IBRs and unexpected tripping caused by low instantaneous frequencies. In this paper, we present a unified optimization framework that bridges frequency stability assessment and UC \& ED studies. The proposed framework works by eliminating solutions that result in instantaneous nodal frequencies that violate a given frequency or RoCoF threshold. Our results indicate that the proposed optimization framework can resolve undesirable, low instantaneous frequencies and high RoCoFs while maintaining reasonable production costs. Future work will focus on incorporating voltage stability as additional constraints.
%


\end{document}